\documentstyle[sprocl]{article}

\newcommand{\nub}{\overline{\nu}}

\catcode`\@=11\relax
\newwrite\@unused
\def\typeout#1{{\let\protect\string\immediate\write\@unused{#1}}}
\def\figurepath{[]}

\def\@nnil{\@nil}
\def\@empty{}
\def\@psdonoop#1\@@#2#3{}
\def\@psdo#1:=#2\do#3{\edef\@psdotmp{#2}\ifx\@psdotmp\@empty \else
    \expandafter\@psdoloop#2,\@nil,\@nil\@@#1{#3}\fi}
\def\@psdoloop#1,#2,#3\@@#4#5{\def#4{#1}\ifx #4\@nnil \else
       #5\def#4{#2}\ifx #4\@nnil \else#5\@ipsdoloop #3\@@#4{#5}\fi\fi}
\def\@ipsdoloop#1,#2\@@#3#4{\def#3{#1}\ifx #3\@nnil 
       \let\@nextwhile=\@psdonoop \else
      #4\relax\let\@nextwhile=\@ipsdoloop\fi\@nextwhile#2\@@#3{#4}}
\def\@tpsdo#1:=#2\do#3{\xdef\@psdotmp{#2}\ifx\@psdotmp\@empty \else
    \@tpsdoloop#2\@nil\@nil\@@#1{#3}\fi}
\def\@tpsdoloop#1#2\@@#3#4{\def#3{#1}\ifx #3\@nnil 
       \let\@nextwhile=\@psdonoop \else
      #4\relax\let\@nextwhile=\@tpsdoloop\fi\@nextwhile#2\@@#3{#4}}
\def\psdraft{
	\def\@psdraft{0}
}
\def\psfull{
	\def\@psdraft{100}
}
\psfull
\newif\if@prologfile
\newif\if@postlogfile
\newif\if@noisy
\def\pssilent{
	\@noisyfalse
}
\def\psnoisy{
	\@noisytrue
}
\psnoisy
\newif\if@bbllx
\newif\if@bblly
\newif\if@bburx
\newif\if@bbury
\newif\if@height
\newif\if@width
\newif\if@rheight
\newif\if@rwidth
\newif\if@clip
\newif\if@verbose
\def\@p@@sclip#1{\@cliptrue}

\def\@p@@sfile#1{\def\@p@sfile{null}%
	        \openin1=#1
		\ifeof1\closein1%
		       \openin1=\figurepath#1
			\ifeof1\typeout{Error, File #1 not found}
			\else\closein1
			    \edef\@p@sfile{\figurepath#1}%
                        \fi%
		 \else\closein1%
		       \def\@p@sfile{#1}%
		 \fi}
\def\@p@@sfigure#1{\def\@p@sfile{null}%
	        \openin1=#1
		\ifeof1\closein1%
		       \openin1=\figurepath#1
			\ifeof1\typeout{Error, File #1 not found}
			\else\closein1
			    \def\@p@sfile{\figurepath#1}%
                        \fi%
		 \else\closein1%
		       \def\@p@sfile{#1}%
		 \fi}

\def\@p@@sbbllx#1{
		\@bbllxtrue
		\dimen100=#1
		\edef\@p@sbbllx{\number\dimen100}
}
\def\@p@@sbblly#1{
		\@bbllytrue
		\dimen100=#1
		\edef\@p@sbblly{\number\dimen100}
}
\def\@p@@sbburx#1{
		\@bburxtrue
		\dimen100=#1
		\edef\@p@sbburx{\number\dimen100}
}
\def\@p@@sbbury#1{
		\@bburytrue
		\dimen100=#1
		\edef\@p@sbbury{\number\dimen100}
}
\def\@p@@sheight#1{
		\@heighttrue
		\dimen100=#1
   		\edef\@p@sheight{\number\dimen100}
}
\def\@p@@swidth#1{
		\@widthtrue
		\dimen100=#1
		\edef\@p@swidth{\number\dimen100}
}
\def\@p@@srheight#1{
		\@rheighttrue
		\dimen100=#1
		\edef\@p@srheight{\number\dimen100}
}
\def\@p@@srwidth#1{
		\@rwidthtrue
		\dimen100=#1
		\edef\@p@srwidth{\number\dimen100}
}
\def\@p@@ssilent#1{ 
		\@verbosefalse
}
\def\@p@@sprolog#1{\@prologfiletrue\def\@prologfileval{#1}}
\def\@p@@spostlog#1{\@postlogfiletrue\def\@postlogfileval{#1}}
\def\@cs@name#1{\csname #1\endcsname}
\def\@setparms#1=#2,{\@cs@name{@p@@s#1}{#2}}
\def\ps@init@parms{
		\@bbllxfalse \@bbllyfalse
		\@bburxfalse \@bburyfalse
		\@heightfalse \@widthfalse
		\@rheightfalse \@rwidthfalse
		\def\@p@sbbllx{}\def\@p@sbblly{}
		\def\@p@sbburx{}\def\@p@sbbury{}
		\def\@p@sheight{}\def\@p@swidth{}
		\def\@p@srheight{}\def\@p@srwidth{}
		\def\@p@sfile{}
		\def\@p@scost{10}
		\def\@sc{}
		\@prologfilefalse
		\@postlogfilefalse
		\@clipfalse
		\if@noisy
			\@verbosetrue
		\else
			\@verbosefalse
		\fi
}
\def\parse@ps@parms#1{
	 	\@psdo\@psfiga:=#1\do
		   {\expandafter\@setparms\@psfiga,}}
\newif\ifno@bb
\newif\ifnot@eof
\newread\ps@stream
\def\bb@missing{
	\if@verbose{
		\typeout{psfig: searching \@p@sfile \space  for bounding box}
	}\fi
	\openin\ps@stream=\@p@sfile
	\no@bbtrue
	\not@eoftrue
	\catcode`\%=12
	\loop
		\read\ps@stream to \line@in
		\global\toks200=\expandafter{\line@in}
		\ifeof\ps@stream \not@eoffalse \fi
		\@bbtest{\toks200}
		\if@bbmatch\not@eoffalse\expandafter\bb@cull\the\toks200\fi
	\ifnot@eof \repeat
	\catcode`\%=14
}	
\catcode`\%=12
\newif\if@bbmatch
\def\@bbtest#1{\expandafter\@a@\the#1
\long\def\@a@#1
\long\def\bb@cull#1 #2 #3 #4 #5 {
	\dimen100=#2 bp\edef\@p@sbbllx{\number\dimen100}
	\dimen100=#3 bp\edef\@p@sbblly{\number\dimen100}
	\dimen100=#4 bp\edef\@p@sbburx{\number\dimen100}
	\dimen100=#5 bp\edef\@p@sbbury{\number\dimen100}
	\no@bbfalse
}
\catcode`\%=14
\def\compute@bb{
		\no@bbfalse
		\if@bbllx \else \no@bbtrue \fi
		\if@bblly \else \no@bbtrue \fi
		\if@bburx \else \no@bbtrue \fi
		\if@bbury \else \no@bbtrue \fi
		\ifno@bb \bb@missing \fi
		\ifno@bb \typeout{FATAL ERROR: no bb supplied or found}
			\no-bb-error
		\fi
		\count203=\@p@sbburx
		\count204=\@p@sbbury
		\advance\count203 by -\@p@sbbllx
		\advance\count204 by -\@p@sbblly
		\edef\@bbw{\number\count203}
		\edef\@bbh{\number\count204}
}
\def\in@hundreds#1#2#3{\count240=#2 \count241=#3
		     \count100=\count240	
		     \divide\count100 by \count241
		     \count101=\count100
		     \multiply\count101 by \count241
		     \advance\count240 by -\count101
		     \multiply\count240 by 10
		     \count101=\count240	
		     \divide\count101 by \count241
		     \count102=\count101
		     \multiply\count102 by \count241
		     \advance\count240 by -\count102
		     \multiply\count240 by 10
		     \count102=\count240	
		     \divide\count102 by \count241
		     \count200=#1\count205=0
		     \count201=\count200
			\multiply\count201 by \count100
		 	\advance\count205 by \count201
		     \count201=\count200
			\divide\count201 by 10
			\multiply\count201 by \count101
			\advance\count205 by \count201
		     \count201=\count200
			\divide\count201 by 100
			\multiply\count201 by \count102
			\advance\count205 by \count201
		     \edef\@result{\number\count205}
}
\def\compute@wfromh{
		\in@hundreds{\@p@sheight}{\@bbw}{\@bbh}
		\edef\@p@swidth{\@result}
}
\def\compute@hfromw{
		\in@hundreds{\@p@swidth}{\@bbh}{\@bbw}
		\edef\@p@sheight{\@result}
}
\def\compute@handw{
		\if@height 
			\if@width
			\else
				\compute@wfromh
			\fi
		\else 
			\if@width
				\compute@hfromw
			\else
				\edef\@p@sheight{\@bbh}
				\edef\@p@swidth{\@bbw}
			\fi
		\fi
}
\def\compute@resv{
		\if@rheight \else \edef\@p@srheight{\@p@sheight} \fi
		\if@rwidth \else \edef\@p@srwidth{\@p@swidth} \fi
}
\def\compute@sizes{
	\compute@bb
	\compute@handw
	\compute@resv
}
\def\psfig#1{\vbox {
	\ps@init@parms
	\parse@ps@parms{#1}
	\compute@sizes
	\ifnum\@p@scost<\@psdraft{
		\if@verbose{
			\typeout{psfig: including \@p@sfile \space }
		}\fi
		\special{ps::[begin] 	\@p@swidth \space \@p@sheight \space
				\@p@sbbllx \space \@p@sbblly \space
				\@p@sbburx \space \@p@sbbury \space
				startTexFig \space }
		\if@clip{
			\if@verbose{
				\typeout{(clip)}
			}\fi
			\special{ps:: doclip \space }
		}\fi
		\if@prologfile
		    \special{ps: plotfile \@prologfileval \space } \fi
		\special{ps: plotfile \@p@sfile \space }
		\if@postlogfile
		    \special{ps: plotfile \@postlogfileval \space } \fi
		\special{ps::[end] endTexFig \space }
		\vbox to \@p@srheight true sp{
			\hbox to \@p@srwidth true sp{
				\hss
			}
		\vss
		}
	}\else{
		\vbox to \@p@srheight true sp{
		\vss
			\hbox to \@p@srwidth true sp{
				\hss
				\if@verbose{
					\@p@sfile
				}\fi
				\hss
			}
		\vss
		}
	}\fi
}}
\def\psglobal{\typeout{psfig: PSGLOBAL is OBSOLETE; use psprint -m instead}}
\catcode`\@=12\relax

\bibliographystyle{99} 
\arraycolsep1.5pt

\def\Journal#1#2#3#4{{#1} {\bf #2}, #3 (#4)}

\def\NCA{\em Nuovo Cimento}
\def\NIM{\em Nucl. Instrum. Methods}
\def\NIMA{{\em Nucl. Instrum. Methods} A}
\def\NPB{{\em Nucl. Phys.} B}
\def\PLB{{\em Phys. Lett.}  B}
\def\PRL{\em Phys. Rev. Lett.}
\def\PRD{{\em Phys. Rev.} D}
\def\ZPC{{\em Z. Phys.} C}

\def\st{\scriptstyle}
\def\sst{\scriptscriptstyle}
\def\mco{\multicolumn}
\def\epp{\epsilon^{\prime}}
\def\vep{\varepsilon}
\def\ra{\rightarrow}
\def\ppg{\pi^+\pi^-\gamma}
\def\vp{{\bf p}}
\def\ko{K^0}
\def\kb{\bar{K^0}}
\def\al{\alpha}
\def\ab{\bar{\alpha}}
\def\be{\begin{equation}}
\def\ee{\end{equation}}
\def\bea{\begin{eqnarray}}
\def\eea{\end{eqnarray}}
\def\CPbar{\hbox{{\rm CP}\hskip-1.80em{/}}}

\begin{document}

\title{DIS PROSPECTS AT THE FUTURE MUON COLLIDER FACILITY}

\author{JAEHOON YU \\
(for the Muon Collider Collaboration \& DIS working group
at the FMC workshop)}

\address{MS309, FNAL, P.O. Box 500, Batavia,
\\ IL 60510, USA\\E-mail: yu@fnal.gov}


\maketitle\abstracts{
We discuss prospects of deep inelastic scattering physics capabilities
        at the future muon collider facility.   
In addition to $\mu^{+}\mu^{-}$ collider itself, the facility provides 
        other possibilities.
Among the possibilities, we present muon-proton collider and 
         neutrino fixed target programs at the muon collider facility.
This $\mu-p$ collider program extends kinematic reach and luminosity 
        by an order of magnitude, increasing the possibility of search for
        new exotic particles.
Perhaps most intriguing DIS prospects come from utilizing high intensity
         neutrino beam resulting from continuous decays of muons in various 
         sections of the muon collider facility.
One of the most interesting findings is a precision measurement of electroweak
         mixing angle, $sin^{2}\theta_{W}$, which can be achieved to 
         the precision equivalent to $\delta M_{W}\sim 30$MeV.
}
\section{Introduction}
Lepton colliders have been used, in general, for precision measurements.
High Energy physics field has been performing precision measurements using 
        electron-position ($e^{-}e^{+}$) colliders due to the fact 
        that electrons do not decay in the acceleration process.
However, $e^{-}e^{+}$ colliders have difficulties in increasing 
        center of mass energy, $\sqrt{s}$, due to synchrotron radiation
        energy losses of electrons (positrons) which is proportional
        to $\frac{1}{m_{e}^{4}}$.
Thus in order to increase the center of mass energy of an $e^{-}e^{+}$ 
        collider one needs a larger physical size of the collider ring.
Since muons, on the other hand, have mass of 200 times bigger than electrons, 
        the synchrotron radiation in a $\mu^{+}\mu^{-}$ collider is $10^{-10}$ 
        less than an $e^{-}e^{+}$ collider, enabling higher $\sqrt{s}$ 
        with the same ring size.

Less radiation energy loss reduces beam momentum spreads and
        allows precision measurements of any possible resonance states
        at the given $\sqrt{s}$.
In addition, since the Higgs coupling is a Yukawa coupling that is proportional
        to the square of the mass of the initial state 
        particles~\cite{th:higgs}, the Higgs production cross
        section at the muon collider is a factor $10^{4}$ bigger than that
        from $e^{+}e^{-}$ colliders at the same $\sqrt{s}$, above the Higgs
        mass threshold.

The baseline strategy for the muon collider complex is as follows :
1) Start out with intense 16GeV proton sources with the beam intensity of
       $6\times 10^{20}$ protons/year eventually increasing to 
       $1.5\times10^{22}$ protons/year~\cite{ex:mucol_acce}.
2) These high intensity protons produce extremely intense low energy muons and
       provide $8\times10^{19}$ muons/year, reaching up to $2\times 10^{21}$
       muons/year.

\begin{figure}[tbp]
\begin{center}
\centerline{\psfig{figure=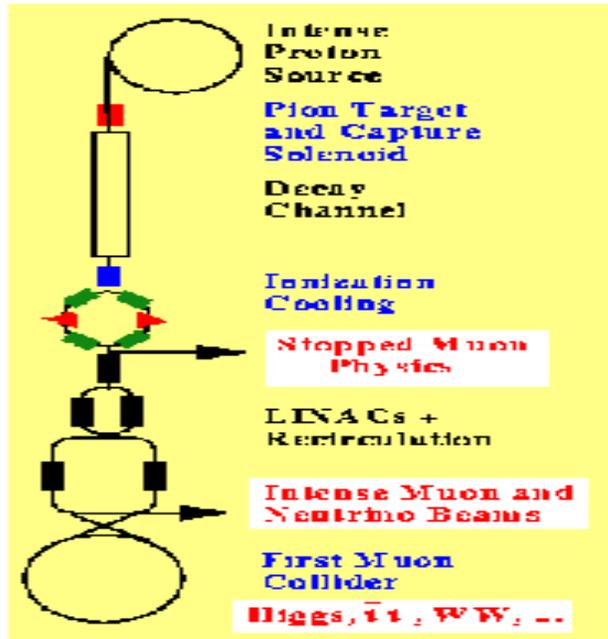,width=3.5in,height=3.5in}}
\caption[]{A schematic view of a muon collider facility.}
\label{fg:schematics}
\end{center}
\end{figure}

Figure~\ref{fg:schematics} shows a schematic view of a muon collider facility.
The discussions presented in this paper consider the following two 
       options :
1) $\mu-p$ collider program with 200GeV muon beam from the muon 
       collider on 1TeV protons from Tevatron.
2) neutrino fixed target program using extremely intense neutrino 
       beam resulting from the decays of 250GeV muons.

\section{Muon-Proton ($\mu-p$) Collider Option}\label{subsec:mup-col}
\begin{table}[t]
\caption[]{Comparison of $\mu p$ collider machine parameters to HERA.}
\label{tb:mup-param}
\begin{center}
\begin{tabular}{|c|c|c|}\hline\hline
 & FMC & HERA \\ \hline\hline
 $E_{l}$ & 200GeV $\mu$ & 27.5 GeV $e^{-}(e^{+})$ \\
 $E_{p}$ & 1000GeV & 800GeV \\
 $Q^{2}$ &$\sim8\times10^5 {\rm GeV^{2}}$ & $\sim9\times10^4 {\rm GeV^{2}}$ \\
 $\int{\cal L}dt$ &10$fb^{-1}/yr$ & 1$fb^{-1}/{\rm life time}$ \\\hline\hline
\end{tabular}
\end{center}
\end{table}
Charged lepton-proton colliders have been used for classical
        deep inelastic scattering experiments, providing ideal
        means of probing nucleon structure.
Thus one of the options in the $\mu^{+}\mu^{-}$ collider complex is
        the $\mu-p$ collider program using 200GeV $\mu$ extracted from the 
        muon collider on 1000GeV protons from the Tevatron.
Table~\ref{tb:mup-param} compares this $\mu-p$ collider machine 
        parameters~\cite{ex:mup-param} to those of HERA's, including the HERA II.
As can be seen in the table, significant improvements can be
        obtained in kinematic reach and integrated luminosity, increasing 
        both by a factor of 10.
\section{DIS in $\mu-p$ Collider}
\begin{figure}[hbp]
\vskip 0.3in
\centerline{\psfig{figure=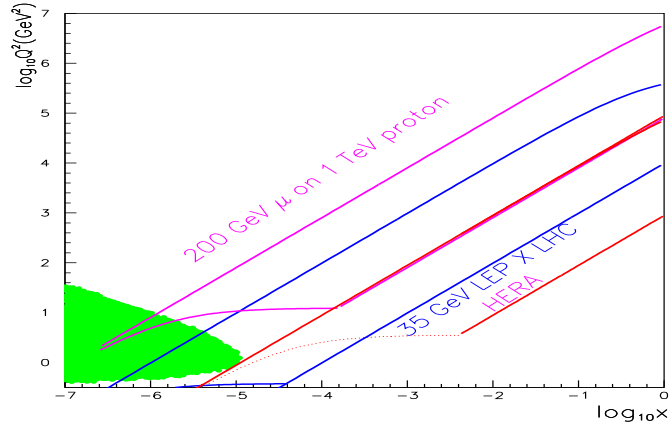,width=4.0in,height=2.5in}}
\vskip -0.75in
\caption[]{The kinematic reach for an asymmetric $\mu-p$ machine.  No 
 $\mu$ collider specific backgrounds have been considered. The nominal
 $y$ and $\theta_\mu$ cuts which correspond to HERA have been applied.}
  \label{fg:mpcol1}
\end{figure}

Most significant improvement one would expect in $\mu-p$ collider option 
         comes from increased kinematic reach and the integrated luminosity.
Figure~\ref{fg:mpcol1} shows the kinematic reach in $log_{10}(Q^2)$ vs
         $log_{10}(x)$ plane, demonstrating the increased $Q^{2}$ and $x$ 
         reach by about a factor of 10 relative to HERA.
This extended kinematic reach in $x$ would enable probing gluon saturation
         regions, indicated as a shaded area in the figure, in $x<10^{-5}$ 
         where the BFKL dynamics would play a role.
However, reaching to this very low-$x$ would require a detector coverage 
         down to a very small angle ($\theta=179^{o}$).
This extreme small angle coverage 
is the biggest hurdle in the $\mu p$ collider
         due to the electron background 
from the constant decay of muons in the
         ring and the finite physical size of the beam pipe.

The extended reach in $Q^{2}$, due to higher $\sqrt{s}$ and a factor 
         of 10 increase in luminosity, would result 
         in higher statistics up to
         $Q^{2}=10^{5}$ regions, allowing improved sensitivity in exotic 
         particle searches.
The configuration given in Table~\ref{tb:mup-param} results in $\sim$1 million
         events per year with $Q^{2}>5000{\rm GeV^{2}}$, compared to 
         326 events from ZEUS experiment~\cite{ex:zeus} in this region with
         $\int {\cal L}dt=34pb^{-1}$.
This high $Q^{2}$ reach also increases the sensitivity for scalar lepto-quark
         to $M_{LQ}=800$GeV to $3\times 10^{-2}$, assuming $\sqrt{s}=1$TeV and
         $\int {\cal L}dt=10fb^{-1}$.
\vskip -0.75in
\section{Neutrino Beam From Muon Colliders}
Perhaps, most intriguing byproduct of the muon collider is very intense 
         neutrino
         beam ($\nu_{\mu}$ and $\overline{\nu}_{e}$) from the continuous
         decays of muons in the straight sections of the muon collider 
         facility~\cite{ex:neu-steve}.
One can expect the neutrino beam in a small straight sections in two 
         components of the muon collider complex that can be seen
         in Fig.~\ref{fg:schematics}. 
The two are recirculating linac (RLA3) and the storage ring.

\begin{figure}[tbp]
\centerline{\psfig{figure=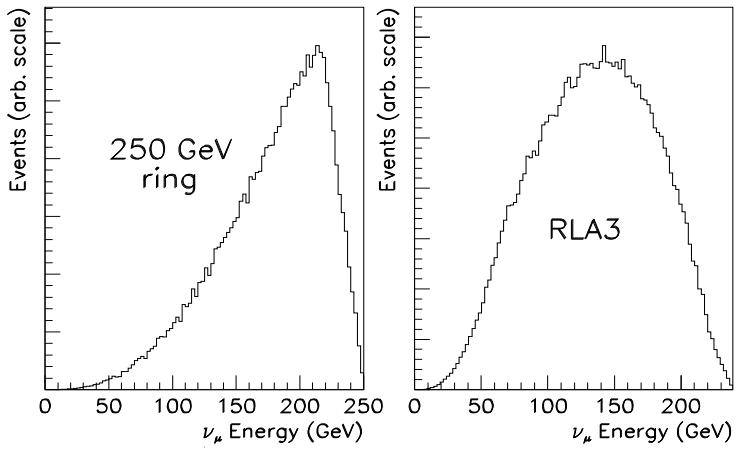,width=4.0in,height=2.5in}}
\caption[]{Energy spectra of neutrino beam from 10m straight section of
250GeV ring (left) and RLA3 (right).} 
\label{fg:enu}
\end{figure}

Figure~\ref{fg:enu} shows the neutrino energy spectra from these two sections
         of the muon collider~\cite{ex:dah-ksm}, assuming 10m straight 
         sections.
Since the angular dispersion of the neutrinos is inversely proportional to 
       the Lorentz $\gamma$ factor of muons, the neutrino beam 
       resulting from the two straight sections
       with $E_{\mu}=250$GeV are very well collimated.
Expected neutrino fluxes from these two sections are approximately 
       1000 time that of
       currently most intense neutrino beam seen at the NuTeV experiment.
Due to the high neutrino flux many DIS possibilities lie in the facility.

\begin{figure}[tbp]
\centerline{\psfig{figure=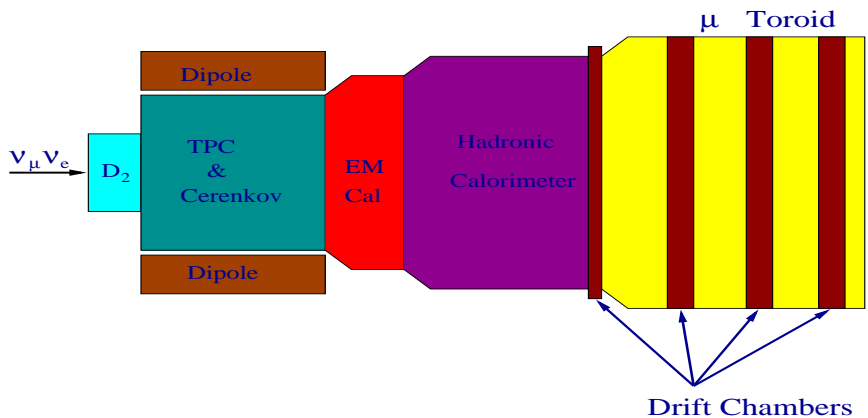,width=4.5in,height=2.5in}}
\caption[]{A light target experiment with a generic neutrino detector.  
 The target in this figure is a 1m thick $D_{2}$ with a few layers of
silicon vertex detector sandwiched inside.}
\label{fg:nu-detector}
\end{figure}
\subsection{Current Neutrino Experiments}
In order to compare the muon collider neutrino program prospects with 
       the current experiments, it is beneficial to review current neutrino
       experiments.
The most common characteristics of the current neutrino experiments is the
       use of heavy target detectors in order to increase statistics, 
       given relatively low neutrino flux.
However, using heavy target introduces undesirable features to physical 
       measurements such as:
1) nuclear effects up to $\sim30\%$ with respect to bare proton target,
2) target dependent isovector corrections,
3) coarse grain calorimetric detectors introduces poor analysis resolution,
and
4) fine grain detectors, in general, have poor energy
       resolution.
Despite the fact that the neutrino experiments have the above disadvantages,
      the neutrino DIS is the only source of measuring $q-\overline{q}$ and
      $V_{dc}$, and provides precision measurement of $sin^{2}\theta_{W}$.

\subsection{A Neutrino Experiment at the Muon Collider Facility}
Unlike the current massive target neutrino experiments, one can imagine 
      an experiment with a light target ($H_{2}$ or $D_{2}$) followed
      by a generic detector that can distinguish electrons from muons resulting
      from charged-current (CC) interactions of $\nu_{\mu}$ or 
      $\overline{\nu}_{e}$.
Since the neutrino flux expected from the muon collider is 1000 times 
      higher than
      current beam, one can obtain sufficient statistics even with light
      neutrino targets.
Figure~\ref{fg:nu-detector} shows a schematic view of such an experiment.
\section{DIS Physics at the Neutrino Beam}
Due to the intensity of the neutrino beam from the decays of
       muons in the straight sections of the muon collider, one 
       would expect large improvements in the current neutrino 
       physics programs.
The first and most immediate measurement one can think of is a precision 
       measurement of nucleon structure function $xF_{3}$ which
       is the only source of $q-\overline{q}$ and is currently statistics 
       limited.
Precise measurement of $xF_{3}$ would provide accurate information on valence
        quarks as well as strange and charm sea quarks.

Heavy quark production measurement is currently statistics limited.
Only di-muon final state is used for heavy quark production in present 
        measurements, due to immediate showering of electrons resulting from
        decays of charmed mesons.
On the other hand, the experiment shown in Fig.~\ref{fg:nu-detector}
        would allow measurements using more, if not all, di-lepton 
        ($\mu\mu$, $ee$, and $e\mu$) final
        states resulting in a total of $\sim 400,000$ di-lepton events 
        per year
        compared to current level of $\sim 5000$ events per entire experiment.
This large increase in statistics would allow accurate measurements of
        CKM matrix elements, $\mid V_{cs}\mid^{2}$ and $\mid V_{dc}\mid^{2}$.

Using a silicon target that is equivalent to $\sim3$m $H_{2}$ would result in
        about 20 events/year of $\mu (e)+b (\overline{b})$ final state assuming
        10m straight section of the collider.
One would be able to increase statistics by lengthening the straight 
        section from 10m to 100m, using thicker target, or running longer.
This increased statistics would allow precision measurement of 
        $\mid V_{ub}\mid^{2}$.
One can also measure neutrino and proton spins, using expected 
        1million events on 200kg target resulting in a 2\% measurement of 
        $\Delta s$.
      
\section{Measurement of $sin^{2}\theta_{W}$}
Measurements of $sin^{2}\theta_{W}$ at the heavy target experiments
        suffer from large statistical uncertainty as well as experimental
        systematic uncertainties in distinguishing CC from neutral
        current (NC) interactions~\cite{ex:ccfr-stw}.
Since the target detectors are heavy materials, the existence of $\nu_{e}$ 
        in the beam causes the current algorithms of distinguishing CC from
        NC using an event length variables to fail, because the CC interaction 
        from $\nu_{e}$ looks identical as the NC events from $\nu_{\mu}$ due to
        immediate showering of electrons.
However, using the experiment shown in Fig.~\ref{fg:nu-detector}, in principle,
        completely eliminates the 
        experimental uncertainties resulting from the definitions of CC and
        NC events.

In addition, since the beam is always a mixture of $\nu_{e}$ and $\nu_{\mu}$,
        CC interactions from both types of neutrinos
        can be used for $sin^{2}\theta_{W}$ measurement, resulting in 
        20million neutrino events in a year from 1m thick $H_{2}$ target.
A preliminary result from NuTeV experiment suffers from statistical 
        uncertainty of the order 1\% based on $\sim 1.5$ million neutrino events
        and major experimental systematic uncertainties 
        from the use of length variable of the order 0.2\%~\cite{ex:nutev-stw}.
Comparatively, one can expect a dramatic reduction in experimental 
        systematic and statistical uncertainties in neutrino experiments in
        the muon collider facility, resulting in a measurement equivalent to
        $\delta M_{W}=30$ MeV~\cite{ex:jae} which is comparable to that 
        expected from
        the TeV33 in the year 2010, using a traditional $M_{T}$ fit.
\section{Technical Challenges}
The most important challenge in the muon collider is a fast and effective
        cooling of the low energy muon beam produced from 8GeV proton 
        interactions on a production target, because the mean life time 
        of muons is 2.2$\mu$sec.
Efficiency of capturing as many muons as possible and of transporting them 
        into acceleration is the most crucial factor in determining 
        the success of a muon collider.
Currently many ideas, including an ionization cooling, have been suggested and
        the muon collider collaboration has recently proposed an ionization 
        cooling experiment~\cite{ex:mucool} at Fermilab.

In addition, there are other challenges to overcome, such as :
efficient focusing, effective shielding of background from $\mu$ decays,
 possible health hazards due to extreme intensity of neutrinos,
extensive and effective beam monitoring, etc.
\section{Conclusions}
Despite many technical challenges listed in the previous section, the 
        muon collider facility
        provides exciting possibilities in DIS physics.
The $\mu-p$ collider option would open up a maximal reach in $Q^{2}$,
        extending the searches of exotic particles.
Reaching to very low $x$ ($x\sim 10^{-5}$) to probe gluon saturation region and
        to study BFKL dynamics is limited due to
        electron background in the beam resulting from muon decays
        as well as an angle limitations of the detectors.
While this 200GeV $\mu$ on 1000GeV proton is the option from the first muon 
	collider, in a longer term future, it is feasible to build a $\mu-p$ 
	collider with 2 TeV $\mu$ on 1 TeV protons with a next generation
	$\mu^{+}\mu^{-}$ collider with $\sqrt{s}=4$TeV.

What is most interesting byproduct of the muon collider is fixed target 
        programs using well understood high flux neutrino beam.
A factor of 1000 increase in neutrino flux relative to present neutrino
        experiments provides wide range of DIS physics possibilities in
        addition to the $\mu^{+}\mu^{-}$ collider itself.

In conclusion, the muon collider is the facility for 
        High-Energy community to open up rich physics capabilities
        in 10 to 15 years down the road.
In order to realize this exciting facility, it is important
        to have constant interest and support within the 
        High-Energy community for the program.

\end{document}